\documentclass{aa}

\usepackage{graphics}

\newcommand{\kms}{km\,s$^{-1}$}
\newcommand{\caii}{\mbox{Ca~{\sc ii}}}
\newcommand{\nai}{\mbox{Na~{\sc i}}}
\newcommand{\feii}{\mbox{Fe~{\sc ii}}}
\newcommand{\mgii}{\mbox{Mg~{\sc ii}}}
\newcommand{\ci}{\mbox{C~{\sc i}}}

\begin{document}
\thesaurus{08 (08.09.2 $\beta\:$Pic; 08.03.4; 08.16.2; 03.13.4)}

\title{
Planetary migration and sources of dust in the $\beta\:$Pictoris disk
}

\author{
A. Lecavelier des Etangs \inst{1}
\thanks{\emph{Present address:}
Institut d'Astrophysique de Paris, 98 Bld Arago, F-75014 Paris, France}
}
\institute{
NCRA, TATA Institute of Fundamental Research, Post Bag 3, Ganeshkhind, 
Pune University Campus, Pune 411 007, India
}

\date{To be published in Astronomy \& Astrophysics}


\maketitle

\begin{abstract}

The dust disk around $\beta\:$Pictoris must be produced by collision or by 
evaporation of orbiting Kuiper belt-like objects. Here we extend the already 
proposed Orbiting-Evaporating-Bodies (OEB) scenario in which the disk is 
a gigantic multi-cometary tail supplied by slowly evaporating bodies
like Chiron. We show that the number of these OEBs must be several
tens of millions, and that this is consistent with the number of bodies
needed to explain the presence of CO and \ci\ in the gaseous disk.

We explore some possible origin of the required perturbation on the OEBs.
If dust is produced by evaporation, a planet with an eccentric orbit can explain the 
observed asymmetry of the disk, because the periastron distribution of the 
parent bodies are then expected to be non-axisymmetric.
Following Malhotra (1995), we investigate the consequence 
for the Kuiper belt-like objects of 
the formation and the migration of an outer planet
like Neptune in Fern\'andez's scheme (Fern\'andez 1982). 
We find that bodies trapped in
resonance with a migrating planet 
can significantly evaporate, producing a $\beta\:$Pictoris-like disk
with similar characteristics like opening angle and asymmetry.

We thus show that the $\beta\:$Pictoris disk can be a transient phenomenon. 
The circumstellar
disks around main sequence stars can be the signature of the present
formation and migration of outer planets.


\end{abstract}

\section{Introduction}
\label{intro}

\subsection{The \object{$\beta\:$Pictoris} disk}
\label{Intro bp disk}

Since their discovery by IRAS in the 80's (Aumann et al. 1984), 
we have now some information about the infrared excess
\object{Vega}-like stars and their circumstellar dusty 
environment. Our know\-ledge 
come essentially from the infrared observations from wh\-ich
can be extracted the spectral energy distribution of the thermal emission 
(Sylvester et al. 1996), which constrains the particle size 
(e.g., Habing et al. 1996) and total dust mass 
(Zuckerman \& Becklin 1993).
Silicate band emissions have also been detected around
some of these stars (Skinner et al. 1992, 
Fajar\-do-Acosta et al. 1993, Knacke et al. 1993,  
Fajardo-Acosta \& Kna\-cke 1995).

However among these infrared excess stars, $\beta\:$Pictoris\ has a very peculiar status
because images have shown that the dust shell is in fact
a disk seen edge-on from the Earth (Smith \& Terrile 1984) and 
have given unique information 
on the dust distribution.
The disk morphology and the inferred spatial distribution of the dust
have been carried out in great details (Artymowicz et al. 1989, 
Kalas \& Jewitt 1995). The morphological properties 
(see Lecavelier des Etangs et al., 1996, hereafter LVF) 
can be summarized as follows:

First, the gradient of the scattered light follows a relatively well-known 
power law.
But the slope of this power law changes at about 120~AU from the star 
(Artymowicz et al. 1990, Goli\-mowski et al. 1993).
The question of whether this change is abrupt or not is still 
open (Mouillet et al. 1997a).
Second, the disk has an inner hole with a central part relatively clear of dust
(Backman et al. 1992, Lagage \& Pantin 1994).
In the third dimension, the disk is a ``wedge'' disk: the thickness increases 
with radius (Backman \& Paresce 1993, Lecavelier des Etangs et al. 1993).
More surprisingly, the disk is not symmetric (Smith \& Terrile 1987):
five different asymmetries have been descri\-bed by Kalas \& Jewitt (1995)
who showed that the disk presents asymmetries in size, brightness and width,
together with butterfly and wing-tilt asymmetries. All these asymmetries except 
the last one show that the disk is not axisymmetric.
An intriguing property is that, despite the asymmetries in brightness
and width, the total brightness integrated perpendicular to the mid-plane 
seems to be, on the contrary, symmetric.
Finally, the inner part of the disk ($\sim 40$~AU) seems to be warped. This
warp has been well-explained by Mouillet et al. (1997b) as due to an inclined
planet inside the disk.

As the dust particle life-time is shorter than the age of the system,
one must consider that the observed dust is continuously resupplied
(Backman \& Paresce 1993). In order to explain the origin
of the dust in the $\beta\:$Pictoris\ disk and these well-known
morphological properties, we have proposed 
the {\em Orbiting-Evaporating-Bodies} model 
(hereafter OEB, see LVF).

After a brief summary of the OEB scenario (Sect.~\ref{OEB et al}), 
and a counting of these OEBs (Sect.~\ref{number of parent bodies}), 
we present plausible origin of the needed perturbation 
on the parent bodies in Sect.~\ref{OEB origin} 
and~\ref{migrating planet}.
We will see that the $\beta\:$Pictoris\ disk can be a natural
consequence of the formation of Neptune-like outer planets.
The conclusion will be found in Sect~\ref{conclusion}.

\section{Summary of the Orbiting-Evaporating-Bodies scenario}
\label{OEB et al}

The observed dust is continuously resupplied.
Two mechanisms can produce dust in this low density disk: collision
or evaporation of kilometer-sized parent bodies (Weissman 1984). 
In both cases, because of the radiation pressure, the 
particles ejected from the parent bodies follow very eccentric orbits
whose eccentricity is related to the grain size (Burns et al. 1979). 
If we assume a zero-order model of a narrow ring of bodies producing dust,
the particles are then distributed on a disk-like structure presenting
three morphological similarities with the $\beta\:$Pictoris\ disk.
First, the central region of the disk is empty of dust,
its limit corresponds to the inner radius of the parent bodies' orbits.
Second, this zero-order model disk is open because the distribution
of the particles inclinations are the same as that of the parent bodies. 
Last, the dust density is decreasing with the distance to the star,
moreover this density distribution follows a power law.
Consequently, if seen edge-on from the Earth, 
the radial brightness profile along the mid-plane of this disk
follows also a power law: $F(r)\propto r^{-\alpha}$ 
(LVF).

We can conclude that a ring of parent bodies on circular orbits
can naturally produce a disk with an inner hole, 
which is open, and if seen edge-on,
the scattered light distribution follows a power law.

But the slope of this power law in the $\beta\:$Pictoris\ disk 
is observed to be $\alpha \sim 4$ (Kalas \& Jewitt 1995).
To explain this distribution with the assumption that
the parent bodies remain in a narrow ring close to the star, 
a large quantity of small particles is needed (LVF).
To solve this, we noticed that if a parent body 
of size $\ga 10$km produces dust by evaporation
and is located at large distances, it produces only small 
particles.
Indeed, if the evaporation rate is small enough,
there is a cut-off on the maximal size of the particles which 
can be ejected from 
the body gravitational field by the evaporating gas.
This slow evaporation and peculiar particle
size distribution is observed in the Solar System around Chiron
(Elliot et al. 1995, Meech et al. 1997). 

In LVF, it has been shown that if some bodies migrate inward from
the outer system, the migration rate is the key
parameter which defines the distribution of the dust produced 
by the evaporation of these bodies.
We find that the migrating rate must be $10^{-5}$ to $10^{-7}$~AU 
per year in order to correspond to the distribution observed around 
$\beta\:$Pictoris\ beyond 100~AU.

Several arguments are in favor of this scenario in the case of 
the $\beta\:$Pictoris\ disk.
First, it is obviously easy to explain any asymmetry even at 
large distances, because a planet in the inner disk (on eccentric orbit) 
can have influence on the distribution of nearby parent bodies, and
this non-axisymmetric distribution is projected outward
by the particle on very eccentric orbits.
Second, this also explains the disk brightness distribution.
This distribution is reported in all published images 
since the first image by Smith \& Terrile (1984) and is 
unsatisfactory left without explanation.
Of course, as shown in~\ref{appendix}, 
the CO/dust ratio is one of the arguments which is in
favour of the OEB scenario.

Finally, the connection between the inner radius of the disk and
evaporation limit is a direct consequence of the OEB scenario because
the periastron distances of the particles are similar to the periastron
of the parent bodies. Any hypothetical planet located at this limit is no
more needed to explain the presence of the inner void in the disk.

\section{The number of Orbiting-Evaporating-Bodies}
\label{number of parent bodies}

Until now, the OEB was described in a qualitative manner.
We do not think that this is a major problem to this model.
For any model of the dust origin, the qualitative aspects 
(dust distribution, morphology, grains properties....) 
and observational prediction are required to validate it. 
Here to become more quantitative, we have just 
to evaluate the number of parent bodies 
corresponding to the production rate inferred from 
the total mass and the dust life-time.
Finally, the last point will be to check that it is a
reasonable value.

\subsection{Bodies producing the observed CO}

Whatever be the origin of the dust, the presence of CO is
undisputed (Vidal-Madjar et al. 1994). 
We do not see any other process than evaporation to
produce it. This gives a strong lower limit on the
number of bodies which are now evaporating.

If we take an evaporation rate of CO: 
$Z_{CO}\sim 10^{19}$~m$^{-2}$s$^{-1}$ (see Fig.~3 of LVF)
from a body with a mean radius $\bar{R} \sim 20$~km
as given by the numerical simulation of the OEB model, this gives
an evaporation rate per body of $Z_{\rm body}=
4\pi\bar{R}^2 Z_{CO}\sim 5\times 10^{28}$body$^{-1}$s$^{-1}$. 
This rate is similar to 
the rate of the CO production observed on the Hale-Bopp comet at more than
6~AU from the sun (Biver et al. 1996, Jewitt et al. 1996) and has been 
qualified as ``enormous'', because it is the same as for the 
comet Halley at 1~AU. Finally, $N_{CO}$ the number of bodies
(with a typical radius of 20~km) now evaporating CO around $\beta\:$Pictoris\ must be
\begin{equation}
N_{CO}=\frac{M_{CO}\tau_{CO}}{Z_{\rm body}\mu_{CO}}
\approx 6 \cdot 10^7 
\label{Nco}
\end{equation}
where $\mu_{CO}$ is the molecular weight, 
$M_{CO}=7\times 10^{20}$kg is the total mass of CO in the disk
(\ref{appendix}),
 and
$\tau_{CO}=2\cdot 10^{-10}$s$^{-1}$ is the photodissociation rate of 
CO (Van Dishoeck \& Black, 1988)
destroyed by the UV interstellar background 
(the stellar extreme ultraviolet flux is very low and negligible).

This number is extremely large but unavoidable 
because CO is {\em observed}. It 
can be larger if the evaporation rate or the typical radii of
the bodies are smaller. Even with the lower limit on the mass 
of CO given in~\ref{appendix}, 
we still have to accept that the number 
of bodies now evaporating around $\beta\:$Pictoris\ and producing the observed CO
must be in the order of tens of millions.

\subsection{Bodies producing the dust}

We can also roughly evaluate $N_{OEB}$, the number of 
bodies necessary to produce the outer {\em dust} 
disk, if this disk is produced by evaporation of CO
like in the OEB scenario. 
This number of OEB is
\begin{equation}
N_{OEB}=\frac{M_d t_d^{-1}}{Z_{\rm body}\mu_{CO}\varphi\phi_{dust/CO}}
\end{equation}
where $M_d$ is the mass of the dust disk, $t_d$ is the dust life-time, 
$\varphi$ is the mass ratio of the dust effectively kept in the disk
to the dust produced, and $\phi_{dust/CO}$ is the mass ratio of
dust to CO. We assume that the disk is about one lunar mass 
($M_d\sim 7\cdot 10^{22}$kg). The dust life time ($t_d=10^4$yr)
is taken from Artymowicz (1997)
\footnote{The dust lifetime used here
is different from the $10^6$ years
assumed in LVF and taken from Backman \& Paresce
1993. The discrepancy comes from a different evaluation of the normal 
optical depth of the disk given by Artymowicz et al. (1989) and Backman et al.
(1992). Anyway, these values are within the uncertainties of this rough 
calculation, and the same result could also be obtained with the life-time 
given by Backman \& Paresce, if $\phi_{dust/CO}$, the mass loading
of the CO gas flow by dust is smaller than assumed here.}.
One can evaluate $\varphi\sim 0.1$. $\phi_{dust/CO}$ is 
very uncertain; we use the recent value given by
Sekanina (1996) on the comet Hale-Bopp observed at more than 
6~AU from the Sun, and we assume that $\phi_{dust/CO}\ga 10$.
Finally we obtain:
\begin{equation}
N_{OEB}\approx 9\cdot 10^7 
\left(\frac{M_d   }{7\cdot 10^{22}{\rm kg}}\right)
\left(\frac{t_d   }{       10^{4 }{\rm yr}}\right)^{-1} \\
\left(\frac{\phi_{dust/CO}}{       10     }\right)^{-1} \\
\end{equation}
We see that, within the uncertainties, this number has the same order 
of magnitude as the number derived above in Eq.~\ref{Nco}. 
As in Eq.~12 of LVF, 
this shows that the mass of dust driven by CO around $\beta\:$Pictoris\ is consistent 
with the total mass of dust in the outer part of the disk. 

\subsection{Conclusion}

These $\sim 10^7$--$10^8$~objects must be compared to the $10^8$--$10^9$
objects believed to be present between 30 and 100~AU from the Sun as 
the source of the Jupiter Family Comets 
(Duncan \& Levison 1997, Morbidelli 1997).
If we consider that these bodies 
have a radius of about 20~km as given by the models presented in 
LVF or given by the models of Sect.~\ref{migrating planet}, 
this corresponds to about an Earth mass. This 
is well below the mass needed to supply the $\beta\:$Pictoris\ disk only 
by collision (30 Earth mass is needed according to Backman et al. 1995).
Evaporation requires less mass of parent bodies, provided that 
some process is able to start the evaporation. As we will
see in Sect~\ref{migrating planet}, a phenomenon believed to have
occurred in the young Solar System can explain this evaporation.

\subsection{$\beta\:$Pictoris\ a transient phenomenon ?}

It could be difficult to imagine that $\sim 10^8$~bodies have
always been active for $\sim 10^8$~years. Only considering the
observed CO gives that during $10^8$~years, 
$M_{CO}\times 10^8 {\rm years}\times \tau_{CO} = 20 M_{\rm Earth}$
of CO must have been evaporated !
It seems unlikely that this large number of bodies have 
been active since the birth of the system.
This gives evidence that either that $\beta\:$Pictoris\ is very young 
(but already on the main sequence, which gives a lower limit 
on the age of $\ga 10^7$~years (Crifo et al., 1997, see also 
discussion in Vidal-Madjar et al., 1998) 
or that $\beta\:$Pictoris\ must be a transient 
phenomenon. There is in fact no reason to believe that the $\beta\:$Pictoris\ 
system was always as dusty as observed today. 
Of course, the idea that this disk is not transient is a consequence of any
model of collisional erosion from asteroid to dust. 
But with other scenarios, we can easily imagine that
a particular phenomenon occurred recently, and that the density 
of the $\beta\:$Pictoris\ disk must be significantly smaller during the quiescent phase
of simple collisional erosion during which the density can be similar to 
the characteristic density of the more common \object{Vega}-like stars.

\section{The origin of the Orbiting-Evaporating-Bodies}
\label{OEB origin}

Since the
OEBs are able to explain some important aspects of the observations,
possible origin of these OEBs, or more exactly the perturbations
necessary to explain their evaporation, have to be explored. 
Indeed, evaporation takes place only when 
a body is formed beyond a vaporization limit of a volatile and is then 
relocated inside this limit. In fact,
the ``relocation'' can have different meaning for an object on a keplerian 
orbit. It can be a simple variation of its semi-major axis or
an increase of its eccentricity. In any case, the most important
is certainly the periastron where most of the evaporation can take place.
In fact, the beginning of evaporation is equivalent to a decrease of the 
periastron distance. 
Here we present some plausible origins of this decrease.
The crucial points will be the possibility to explain the observed dust
distribution, and some properties like the opening angle or the asymmetries.

In all the simulations presented here and in the next section,
we use the mass of $\beta\:$Pictoris: $M_*= 1.8 $M$_{\odot}$ (Crifo et al. 1997).

\subsection{Perturbation by several planets}
\label{random walk}

As already mentioned in LVF, it 
is very surprising that if the OEBs are supposed to migrate 
slowly inward from large distances, the speed of the evolution 
of the OEBs' orbits (decreasing of their semi-major axis) is similar to 
the estimated motion of the Kuiper belt objects
in the Solar System just beyond Neptune between 30 and 40~AU
(Torbett \& Smoluchowski 1990). This chaotic motion is due to the
perturbation by the four giant planets of the Solar System
and mainly Neptune. If several planets are present inside the 
evaporation limit with the last one close to this limit, 
similar motion can occur around $\beta\:$Pictoris. For example, if the last planet was 
at 20~AU, it would excite putative bodies just beyond and supply a $\beta\:$Pictoris-like
disk with CO$_2$ as the main volatile; the same
can be concluded for a planet around 100~AU with CO as the main volatile.

Here the asymmetry can simply be due to an eccentric orbit of
the perturbing planet. For instance, one major planet on an eccentric orbit
can cause a modulation of the precessing rate 
of the periastron of the OEBs. It is thus well-known that the distribution of
the perihelion of the asteroids in the Solar System is not
axisymmetric, and is closely related to the Jupiter longitude
of the perihelion (Kiang 1966). The density of asteroids with the same longitude 
of perihelion as Jupiter is thus $\sim 2.5$~times 
larger than that with periastron in opposite direction. 
This is simply because 
when the periastron of an asteroid is located at 180 degrees 
from the periastron of Jupiter, the precessing rate is quicker  
and the density is smaller.

\begin{figure}[tb]
\resizebox{\hsize}{!}{\includegraphics{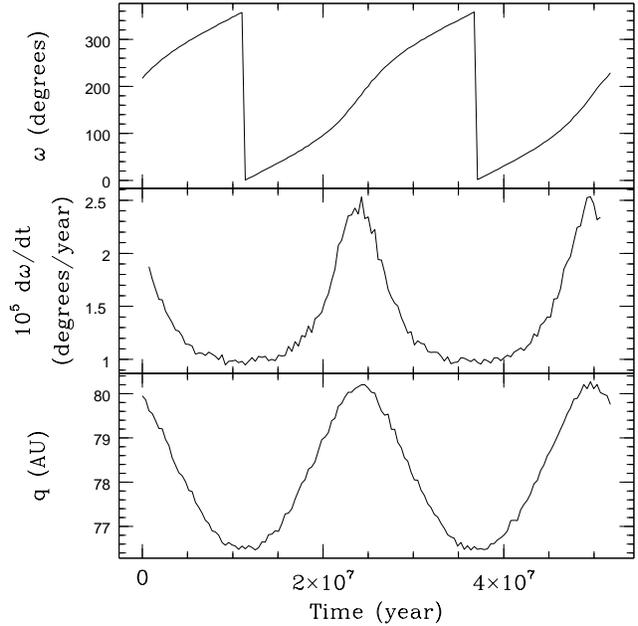}}
\caption[]{
Example of the time evolution of the longitude of periastron ($\tilde\omega$)
of a body perturbed by a massive planet 
on eccentric orbit ($M_p= 3\cdot 10^{-4} $M$_{\odot}= 1.7\cdot 10^{-4} $M$_* $,
$e_p=0.05$, $a_p=30$~AU).
The body is relatively far from the planet with a quasi-constant 
semi-major axis ($a\approx 82$~AU) and oscillating eccentricity ($0.03<e<0.07$).
The precessing rate of the periastron ($d\tilde\omega/dt$) and the periastron distance
($q$) are larger when the longitude of the periastron is at 180 degrees from 
the periastron of the planet
}
\label{dwdt1}
\end{figure}

\begin{figure}[tb]
\resizebox{\hsize}{!}{\includegraphics{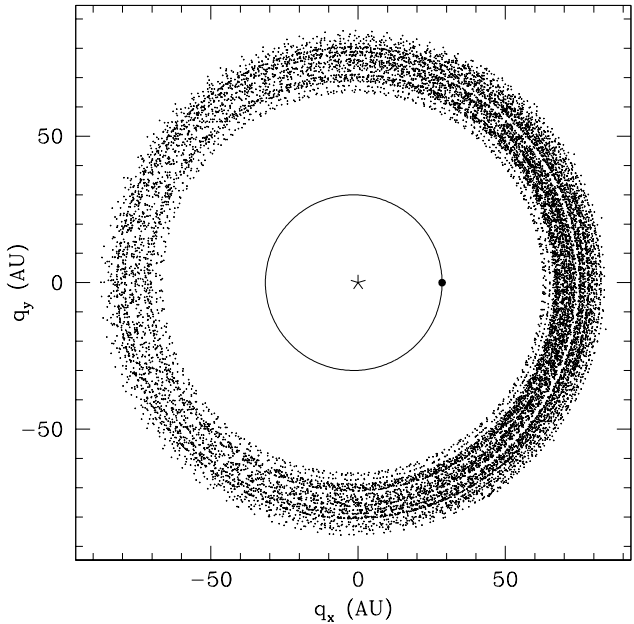}}
\caption[]{Plot of the spatial distribution of the periastron of a set of 
bodies located between 70 and 90~AU and perturbed by a planet with the
same characteristics as in the previous figure.
We see that the density of bodies with periastron in the direction of 
the planet periastron (black dot) is very large.
Moreover the periastron distances are also smaller. For these two reasons, 
the dust production by evaporation must be larger in this direction. 
If these bodies evaporate, they produce a dust disk which must be asymmetric.}
\label{dwdt2}
\end{figure}

Such an effect would obviously
cause an asymmetry in a disk \emph{if it is
produced by evaporation} of bodies with a distribution
of periastron perturbed in this way.
As the dust is mainly produced at the periastron of the parent bodies 
and principally observed during the apoastron, the part of the disk 
at 180 degrees from the
perturbing planet periastron could be more den\-se.

For example, we have evaluated the distribution of the periastron 
of a belt of bodies between 70 and 90~AU perturbed by a massive planet
(mass $M_p= 3\cdot 10^{-4} $M$_{\odot}=1.7\cdot  10^{-4} $M$_*$) 
with an eccentricity $e_p=0.05$ and a semi major-axis of $a_p=30$~AU
(Fig~\ref{dwdt1}). We find that, with bodies' inclinations between
0 and 7 degrees and eccentricities between 0 and 0.12 
($i^2=e^2$, with $i$ in radians), the density of bodies with 
same longitude of periastron as the planet is $\sim$3 times larger
than that in the opposite direction (Fig~\ref{dwdt2}). 
This value of the ratio does not depends on the distance
but is smaller
if the bodies' eccentricities are larger. As a conclusion, 
a planet on an eccentric orbit 
can be responsible for asymmetry in the distribution of the 
orbital parameter of the parent bodies of dust disks, and thus creates 
asymmetry in the observed disk if produced by evaporation
of the parent bodies at periastron.\\

In the scenario of chaotic motion due to 
several planets, the $\beta\:$Pictoris\ disk is not a transient phenomenon but
quiescent, the number of parent bodies already evaporated 
from the beginning of the process must be very large. 
However, catastrophic scenario can also be developed. As we will see,
the $\beta\:$Pictoris\ disk can be a transient phenomenon.\\

\ \\

\subsection{A ``cat-planet''}

For two years, planets have been indirectly discovered through radial 
velocities measurements. These planets have very unexpected properties, 
with giant planets very close to the stars like 51~Peg
(Mayor \& Queloz 1995), or with very high eccentricity like the planet
around 16~Cyg~B with an eccentricity as high as 0.6 (Cochran et al. 1997). 
The origin of this eccentricity is subject to debate.
Several plausible solutions have been proposed like 
high eccentricity induced by the companion (16~Cyg~A) (Mazeh et al. 1997),
or series of planet-planet encounters (Weidenschilling \& Marzari 1997).

In this last scenario, during an encounter between two massive planets, one is
ejected on very eccentric orbit. It should be realized that if 
this planet is then put through a Kuiper-like belt 
as the cat among the pigeons,
this creates a situation where numerous bodies can be ejected 
on evaporating orbits.

This is a catastrophic scenario, which might have happened recently
around $\beta\:$Pictoris.
But in that case some questions arise:
Why is the \object{Vega}-phenomenon so common? Is $\beta$~Pic\-to\-ris 
really different from other \object{Vega}-like stars?
In that context, there could also be a link 
with the numerous Falling-Evaporating-Bodies (FEB),
which are in fact cometary objects
on very eccentric orbits (Vidal-Madjar et al. 1998, see 
also~\ref{appendix}).
But this link remains very weak.

\begin{figure}[tb]
\resizebox{\hsize}{!}{\includegraphics{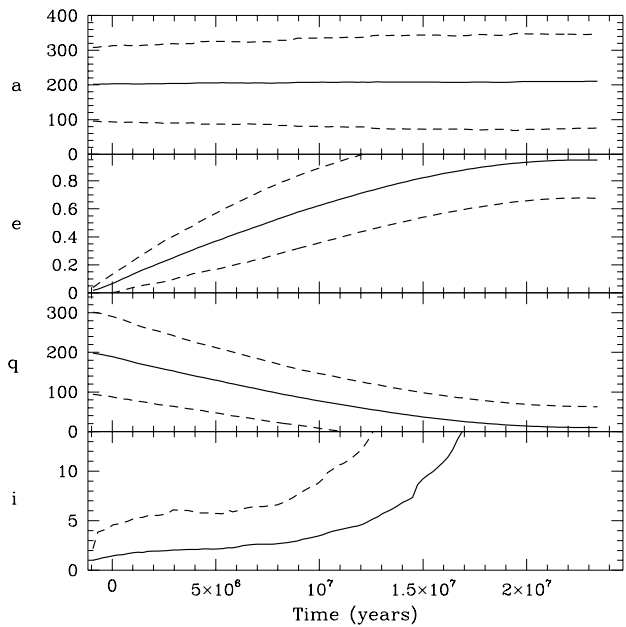}}
\caption[]{Plot of the orbital parameters semi-major axis ($a$), 
eccentricity ($e$), periastron ($q$) and inclination ($i$) of the 
Kuiper-like ``pigeons-bodies'' perturbed by a ``cat-planet''.
Mean values are given (solid lines) with $\pm 2\sigma$ deviations
(dashed lines).
The planet has a mass $M_p=10^{-4}$~M$_{\odot}=5.6 \cdot 10^{-5}$~M$_*$, an eccentricity
$e_p=0.95$ and a semi-major axis $a_p=300$~AU (and
consequently periastron $q_p=15$~AU, and apoastron $Q_p=585$~AU).

The semi-major axis of the ``pigeons-bodies'' remain roughly constant, but 
their eccentricities increase and their periastrons decrease at a rate of 
about $10^5$~AU per year. These bodies can thus produce dust by evaporation.
As expected for the $\beta\:$Pictoris\ disk, they produce only small particles 
because they will be exhausted before the periastron is small enough 
for the gas drag to be able to eject the larger particles.
Note that the inclinations significantly increase to several degrees,
this is the same order as the opening angle of the $\beta\:$Pictoris\ disk.}

\label{cat}
\end{figure}

With these potential objections in mind, we have however 
performed some simple test-simulations to see the behaviour of these
``pigeons-bodies''. Surprisingly, the result is very close
to what is needed to produce a disk by evaporation.
With a planet on a very eccentric orbit 
($e_p\ga 0.95$), and a mass larger than the mass of Neptune, 
{\em all} the Kuiper-like belt objects taken to be randomly distributed between
100 and 300~AU have a slow decrease of there periastron on a time scale of 
\begin{equation}
\tau \sim 10^6 \left( \frac{M_p}{M_J}\right) 
\left( \frac{a_p}{300 {\rm AU}}\right )^{3/2} {\rm years}
\end{equation}
where $M_p$ and $M_J$ are respectively the mass of the planet and Jupiter,
and $a_p$ is the planet semi-major axis (Fig~\ref{cat}).
Because the periastrons of these bodies slowly decrease, these
bodies will produce a disk by evaporation with the required particle 
size distribution and an inner void of matter.
Moreover, simultaneously, there is an increase of the 
inclinations to few degrees as needed to produce an open ("wedge") disk. 
Of course, if produced by the evaporation of these
``pigeons-bodies'', this disk is found to be strongly non-axisymmetric.

Finally, this scenario of the ``cat-planet'' could explain ma\-ny
observed properties of the $\beta\:$Pictoris. But, if true,  
in addition to the fact that this scenario 
still looks very unlikely, we would have to accept that the 
$\beta\:$Pictoris\ disk is completely different by nature from the dust disks 
present around other infrared-excess main sequence stars.

\section{Resonances with a migrating planet.}
\label{migrating planet}
 
We develop now the most promising possible origin for the OEBs.

We propose a possible origin of the OEBs as an extension of the ideas of
Fern\'andez \& Ip (1996), who showed
that the outer planets migrated by several AU during their formation.
As the time of motion is few hundreds of million years, 
it is likely that a similar phenomenon takes place around
the young main sequence stars, around which circumstellar disks
have been observed ($10^{8}$~years is about the age of 
$\beta\:$Pictoris\ and \object{$\alpha$~PsA}). 
Here, we evaluate the dynamical consequence of 
the outer planet migration on the ``planetesimals'', and particularly 
on the evolution of their orbits, the resonance trapping, and 
the possible evaporation.

\subsection{The formation of Uranus and Neptune}
\label{fernandez model}
 
It is now generally accepted that the formation of the Solar System could be
divided into several stages starting with the accretion of Jupiter and Saturn
in the gaseous solar nebula, and then followed by the accumulation of the 
terrestrial planets and the two outer planets, Uranus and Neptune (Lissauer
1993). But the formation of the outer planets of the Solar System raises a lot
of problems in particular the problem of time scales: the planetesimal 
accumulation time scale must be reasonably short, in any case shorter than the
age of the Solar System itself (Lissauer 1987).  
 
Several solutions have been 
proposed to explain the accumulation of a large number
of planetesimals into planets 
at several tens of~AU, where the revolution 
periods are very large. For example, it has been proposed that the planet
embryos could be driven by gas drag.
 
A smart solution has been proposed by Fern\'andez (1982) who suggests that  
the accumulation and scattering of a large number of planetesimals 
is the origin of the migration of the outer planets during their formation.
This migration is essentially due to the exchange of angular momentum 
between Jupi\-ter and the proto-Uranus and proto-Neptune, via the accretion 
and gravitational scattering of planetesimals, the orbit of Jupiter loses 
angular momentum and shifts slightly inward, while those of Saturn, Uranus and
Neptune move outwards, by several~AU for Uranus and 
Neptune. This model successfully explains the formation of the two outer 
planets of the Solar System, in short time scale 
($2\cdot 10^8$ to $3\cdot 10^8$~years),
their mass and their actual position (Fern\'andez \& Ip 1996).
Moreover, in this scenario, the sharp cutoff of planet mass at Neptune's 
distance is due to the orbital expansion of the proto-Neptune, that allowed it
to accumulate bodies from a wide zone of the proto-planetary disk. 
 
The consequences of this scenario on the structure of the outer Solar System 
has been investigated by Malhotra (1993, 1995) who showed that this also
explain the particular orbit of Pluto with large eccentricity and inclination,
and its resonance with Neptune. In short, Pluto is trapped in the 
orbital commensurability moving outward during the expansion phase of 
Neptune's orbit. The consequences on the Kuiper belt have also been analyzed
(Malhotra 1995) and it has been demonstrated that the outward migration of 
Neptune can explain the fact that numerous Kuiper belt objects are observed
in Pluto-like orbit in 2:3 resonance with Neptune.
 
\subsection{Planet migration and perturbation on parent bodies.}
\label{migration}

Here we propose to extend this work to evaluate the link
between the migration of outer planets and the $\beta\:$Pictoris-like circumstellar disks
for which we know that the age is similar to the time scale of formation of
these planets. 
Following Malhotra, we numerically investigate the consequence of the migration
of the planets in the Fern\'andez's scheme
on the dynamical evolution of the planetesimals, and their possible
trapping in resonant orbits which allow evaporation of frozen volatiles.

For simplicity we consider only one outer massive planet 
supposed to suffer an exchange of orbital angular momentum 
as a back-reaction on the planet itself of the planetesimal scattering
in the process of clearing the inner planetary region of residual 
planetesimals. 
Of course, at least a second inner planet must be there, in particular
if the outer is migrating outward. Here, we consider only the principal
outer perturbing planet which is supposed to migrate 
because of a force equivalent to a drag force decreasing 
with time: $F_D \propto e^{-t/\tau}$, where $\tau$ is the characteristic
time of the migration.
We only consider the effect of this migration on the outside bodies.
CO is the only volatile considered for now.

We first look at the evolution of the orbits of a swarm
of planetesimals (test particles). These bodies are 
swept by the resonances migrating simultaneously with the 
planet;  we will check if they are trapped in one of these 
resonances and begin to migrate with it.
Then, we evaluate if the resonant bodies can start 
to evaporate and produce dust.

\subsubsection{Resonance trapping}

In fact, if the migrating planet is moving inward, the planetesimals
are not permanently trapped in the resonances. Their
semi-major axis remain unchanged and their eccentricities are 
only slightly increased. Consequently, the decrease of the periastron
distance is too small to allow the volatiles to evaporate.

On the contrary, if the planet is moving outward, as given by the 
results of Fern\'andez \& Ip (1996) which show an expansion of
the Uranus and Neptune orbits,
a fraction of bodies can be trapped in resonances.
Their semi-major axis and eccentricities increase significantly
and the net result is a decrease of their periastron.
If $q_i$, $q_f$, $a_i$ and $a_f$ are respectively the initial and final 
periastron distances and semi-major axis of the body, we have
$q_f/q_i \sim (1-\sqrt{\ln \frac{a_f}{a_i}}) \frac{a_f}{a_i} $. 
Thus $q_f/q_i <  0.75 $ as soon as  $\frac{a_f}{a_i}> 1.1$.
As soon as a body is trapped in the resonance, the
ratio of the final to the initial semi-major axis is the same
for the body as well as for the planet.
The decrease of the periastron of the body can be significant 
even if the planet's semi-major axis is increased by only ten 
percent. This can start the evaporation of trapped bodies.

\begin{figure}[tb]
\resizebox{\hsize}{!}{\includegraphics{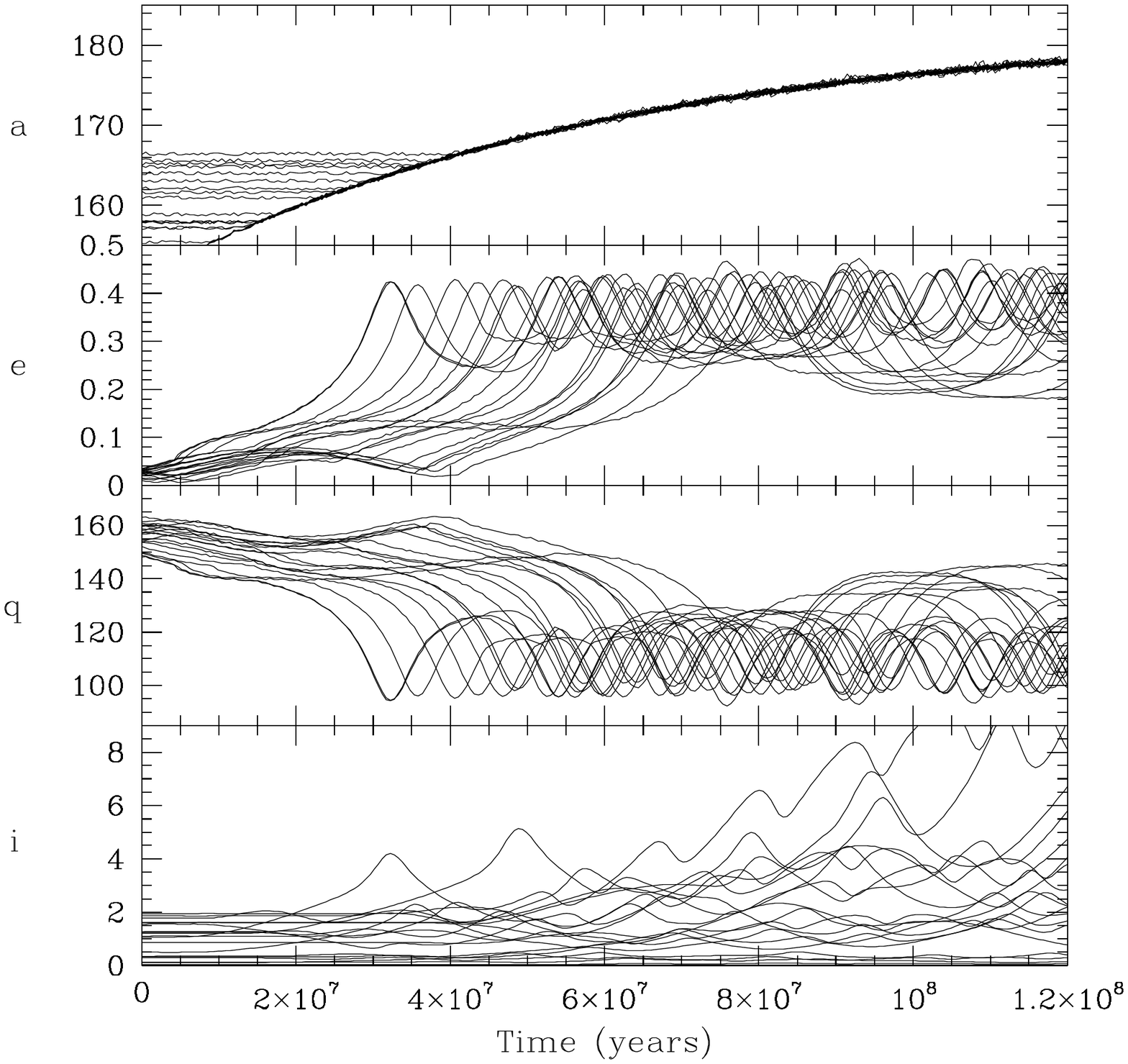}}
\caption[]{Plot of the orbital parameters semi-major axis ($a$), 
eccentricity ($e$), periastron ($q$) and inclination ($i$) of 19 bodies trapped
in the 1:4 resonances with a migrating planet.

We see that although the semi-major axis of the bodies trapped in 1:4 resonances 
increase, their periastron decrease. These bodies can
start to produce dust by evaporation, but because their periastron is still
larger than 100~AU, this evaporation produces only small particles.}
\label{qei2tres4}
\end{figure}

\begin{figure}[tbh]
\resizebox{\hsize}{!}{\includegraphics{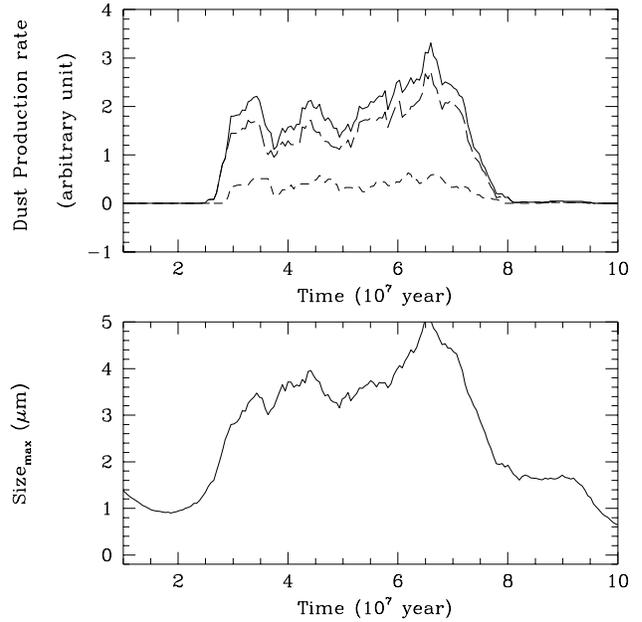}}
\caption[]{Plot of the dust production as a function of time for evaporating 
bodies trapped in 1:4 resonance with a migrating planet.
This is the dust production for grains larger than 2$\mu$m, 
because smaller grains 
are supposed to be quickly expelled by radiation pressure.
Indeed, for particles with radius $s\ge 1\mu$, the ratio 
of the radiation force to the gravitational force ($\beta$)
is roughly inversely proportional to 
the size of the particle, that is to say $\gamma=\beta s$ is constant
(Artymowicz 1988). 
$\gamma$ depends on the grain composition but is 
close to 1$\mu$m  within a factor of two.
Here, $\gamma = 1\mu$m has been assumed.

The production of dust starts when the periastron distance of the
parent bodies is small enough for the CO production rate to
allow ejection of grains larger than 2$\mu$m.
Then, it stops when the parent bodies are exhausted and 
have no more volatile. 
This last condition limits the duration of the phenomenon
and gives a typical radius of evaporating bodies around 20km.

The production rate is also not axisymmetric. As in Sect.~\ref{random walk}, we see 
that the production is 
larger in the direction of the periastron of the perturbing planet (long dashed)
than in the opposite direction (short dashed).

The bottom panel gives the corresponding maximal size of grains ejected 
from the bodies by the evaporating gas.
In this simulation, the maximal particle size is around 4$\mu$m
as expected to explain the slope $\alpha\sim 4$ observed in the $\beta\:$Pictoris\ case. }
\label{pna2t}
\end{figure}

\subsubsection{Evaporation}

We have tested several configurations of outward migration
and have evaluated the effect on planetesimals in the zones swept
by first order resonances. 
We have arbitrarily considered only one planet starting 
from an initial semi-major axis $a_i=60$~AU 
and migrating of 12~AU to a final semi-major axis $a_f=72$~AU ($a_f/a_i=1.2$).
The velocity dispersion of the planetesimals swarm is given
by the inclinations randomly chosen between 0 and 2 degrees, 
and the eccentricities between 0 and 0.035. Realistic changes in
the set of inclinations and eccentricities 
do not change the final result, except
for the fraction of bodies effectively trapped in resonances.

The mass of the planet has been taken between Jupiter and Saturn mass.
A less massive planet gives less efficient trapping.
The eccentricity of the planet allows to test the possible influences on
the asymmetry of the dust disk produced by the bodies trapped
in ``evaporating orbit''. 

The conclusion of several test simulations can be summarized as follows:

- The 1:2 resonance is a very efficient resonance for trapping,
even with a very fast migration ($\tau \sim 10^6$years). But the 
variation of the periastron distance is too small 
to allow the evaporation ($\Delta q=q_f-q_i\approx - 10$~AU).

- The 1:3 resonance is efficient if $\tau \ga 10^7$years and can give
a large decrease of the periastron ($\Delta q\approx -30$ -- $-40$~AU).
But with a planet eccentricity of 0.05, no azimuthal asymmetry
has been observed on the distribution of the periastron of the
trapped bodies. Finally, there is no change in the distribution of 
the inclination which remain low.

- The 1:4 resonance give interesting results presented in 
Fig.~\ref{qei2tres4} and~\ref{pna2t}. The trapping has been found to 
be efficient with rather extreme parameters: the mass of the planet
must be $M_p \ga 0.5 M_J$, where $M_J$ is the mass of Jupiter, 
the migration rate must be low 
($\tau \ga 5\cdot 10^{7}$~years). The eccentricity of the planet 
has been taken to be 0.05. With these conditions, 19/500 bodies 
have been found to be trapped in the 1:4 resonance, and the periastron
decreased by $\Delta q\approx -40$~AU. A significant increase 
of the inclination has been observed after few $\tau$ as well as 
a large asymmetry in the distribution of the periastrons longitude.

In evaluating the evaporation rate of this configuration, 
we find that the periastron decrease due to the 1:4 resonance
pre\-sents two favorable particularities.

\begin{enumerate} 

\item This decrease is large enough to start the evaporation of big bodies 
and the ejection of large particles 
which are not too sensitive to the radiation pressure and remains 
on elliptical orbits
(the ratio of the radiation force to the gravitational force
must be $\beta<0.5$).
Indeed, the evaporation must be strong to produce ``$\alpha$-particles''
which are larger than the submicronic ``$\beta$-particles'' ejected 
on hyperbolic orbits
by the radiation pressure, and to extract
these $\alpha$-particles from the gravitational field of parent bodies 
which are large enough to have still volatiles.

\item This decrease of the periastron is also not too large 
and the evaporation rate is low as expected to produce 
small particles needed to explain the observed slope of 
the radial brightness profile (Fig.~\ref{pna2t}).
The evaporation and associated dust production rate 
are shown in Fig.~\ref{pna2t}.

\end{enumerate}

- The 1:5 resonance is efficient in trapping
only if the parameters of the migrating planet are extreme with 
$M_p\ga M_J$, $e_p \ga 0.1$ and $\tau \ga 5\cdot 10^{7}$~years.
With these conditions, we find 10/600 bodies trapped in
the resonance. The consequence on the CO-evaporation rate 
as a function of time is very similar to the one found 
in the 1:4 resonance.
Among the 10 bodies trapped in the 
resonance one has evolved in a very eccentric orbit with large
inclination.
The link with the FEBs is possible but still to be investigated.

\subsection{Time evolution of the disk density}

As the dust life-time is very short ($t_d\sim 10^4$~yr in the $\beta\:$Pictoris\ disk
but also $t_d\sim 10^6$~yr around \object{$\alpha$~PsA}) and shorter
than the stellar ages, the density of dust observed around
the main sequence stars is 
directly related to the actual production rate of dust.

The time variation of the dust production rate, and consequently 
of the disk density,
have been evaluated in the simulations described in Sect.~\ref{migration}.
Fig.~\ref{pna2t} shows the production as the function of time  
for the 1:4 resonance.
The production of dust occurs between $3\cdot 10^{7}$ and $7\cdot 10^{7}$~years
which is very consistent with the estimated age of $\beta\:$Pictoris\ (Crifo et al. 1997).
We also conclude that the dust production rate by evaporation is transient
and can be large during phase during which the bodies trapped in the 
resonances are entering in the evaporation limit. 
The $\beta\:$Pictoris\ disk can be in such a state 
while other \object{Vega}-like stars are in quiescent phase.

\subsection{Asymmetry}

If the bodies are trapped in a resonance with a planet on eccentric
orbit, there can be an asymmetry in the distribution of the 
periastron as already seen in Sect.~\ref{random walk}. 
For example, the Fig.~\ref{pna2t} gives the dust production rate by
the bodies trapped in the 1:4 resonance with a planet on an eccentric
orbit ($e_p=0.05$). The production rate is larger 
in the direction of the periastron of the planet
than in the opposite direction.
The disk thus produced must be asymmetric with a larger density
in the direction of the apoastron of the migrating planet.

\subsection{Opening angle}

From the comparison of Fig.~\ref{qei2tres4} and~\ref{pna2t}, we
can conclude that, in this configuration, the production
of dust takes place between $3\cdot 10^{7}$ and $7\cdot 10^{7}$~years
when the inclination of the parent bodies are still low,
but already larger than the initial inclination ($<2^\circ$).
Here, the opening angle of the produced disk
must be around 3 or 4 degrees.

This angle is smaller than the 7 degrees observed in the $\beta\:$Pictoris\ case.
However, we must remember that we have taken only one planet, 
the ascending node of its orbit is then constant. 
But in a more realistic case with several giant planets, the precession
of the ascending nodes can produce a significant increase of
the parent bodies inclination. It is reasonable to believe that
the process which gives the large inclination
of Pluto or Kuiper belt objects in Malhotra's simulations with 
several planets (Malhotra 1995) can give a large inclination of 
the parent bodies in the $\beta\:$Pictoris\ disk. 
This migrating and resonance 
trapping process can give large increase in the inclinations
up to several degrees (bottom panel of Fig.~\ref{qei2tres4}), and
consequently give a large opening angle of the associated dust disk.

\subsection{Stability of trapping}

We have checked that the bodies trapped in the far resonance 1:4 with the
parameters given above are
steadily trapped in spite of some perturbations expected to be there. 

For example, the motion of the perturbing planet
resulting from the scattering of large planetesimals could cause 
a shift of the resonance and a disruption of the resonance trapping.
However, simulations show that 
the perturbation has a negligible effect as soon as $\Delta \vec{v} / \vec{v} \la 10^{-3}$.

The resonance stability to non-gravitational forces acting on evaporating
bodies has also been checked. For non-gravita\-tio\-nal forces in the 
form $\vec{F}=-A_1 \vec{F_g}$ where $\vec{F_g}$ is the central star gravitation
and $A_1 \la 500\cdot 10^{-5}$ (Sekanina 1981), no consequence
on the resonance trapping has been observed.

Perturbation from other planets will be evaluated in a forthcoming paper. 
Some effects can be expected like the increase of the bodies inclination.
Preliminary results show that the effect on 
the stability of the trapping and the statistics of the number of bodies
which are trapped is negligible.

\subsection{Conclusion}

The consequences of the Fern\'andez's scenario on 
the small bodies of our own Solar System are limited
to dynamical phenomena like the orbits of Pluto or 
the structure of the Kuiper belt
as proposed by Malhotra, because the distance of
Neptune is large in comparison to the evaporation limit of the 
common volatiles. The situation can be different around brighter
stars like $\beta\:$Pictoris\ or stars surrounded by smaller planetary
systems.

We have shown that the formation and migration 
of one outer planet in the Fern\'andez's scheme can explain
the observed properties of the dust disk around $\beta\:$Pictoris.
Indeed, it has been shown that planetesimals can be trapped in the resonances
with the planet in such a way that
it is likely that they produce dust and gas by evaporation.
 
If this model is confirmed to be successful and the evolved disks are 
the signatures of current dynamic evolution around main sequence stars, 
future
observations of circumstellar disks will give information 
on the structure and the time scales 
of the formation of the outer planets in planetary systems.

\section{Conclusion}
\label{conclusion}

Collisions and evaporation are the two main processes believed to be able to
supply disks like the $\beta\:$Pictoris\ one. These two processes 
are not exclusive:
both can occur at different places or with different time-scales.

However, the $\beta\:$Pictoris\ disk is more likely produced by
the evaporation process. 
The CO and \ci\ gas detected
with HST definitely shows that evaporation takes place around $\beta\:$Pictoris, even
if its 
consequence on dust replenishment 
in comparison to the collisional production is still a matter of debate.
The dust spatial distribution with the slope of the power 
law, and the central hole can be explained 
by the characteristic distances of evaporation. Finally, 
the asymmetry at large distances can easily be explained 
in evaporating scenarios because
the parent bodies are maintained close to the star 
where planets' influences are important. The asymmetry 
is then simply a consequence of the non-axisymmetry of the perturbation 
by planet(s) on eccentric orbits.

We have presented some possible origins of the OEBs. The direct consequence
of some mechanisms to drive the parent bodies inside the evaporation zone
is that the $\beta\:$Pictoris\ phenomenon can be transient, 
with periods of large activity.

We have shown the possibility that bodies trapped in resonances with a 
migrating planet can evaporate. This allows to explain the
asymmetry observed in the $\beta\:$Pictoris\ disk; the needed large number of CO
evaporating bodies is realistic and explained by
a transient evaporation during a short period.

From another point of view, if the migration of the outer planets 
took place in the Solar System and had
consequences on small bodies like Pluto or the Kuiper belt objects,
why not around other stars ?
This is in fact a simple consequence of the presence of a forming
planet inside a disk of residual planetesimals: 
the interaction between the planet and these planetesimals can eject
some planetesimals on eccentric orbits (FEBs?) and this ejection
causes a back-reaction on the planet which migrates. In details,
as shown by Fern\'andez (1982), several planets
are needed and the planetesimals are the medium for the momentum exchange
between these planets.
Here we have explored the new consequence of the migration of a forming planet.
The small bodies can be trapped in resonances, and then, 
because their eccentricities increase, they can enter inside
evaporation zone and become parent bodies 
of $\beta\:$Pictoris-like disks. In short, as a direct consequence of the formation
of outer planets in the Fern\'andez's scheme, evaporation of Kuiper belt-like
objects around stars can be expected to be common. 
This allows us to look at the circumstellar disks around main sequence 
stars as a possible signature of outer planet formation. 

\begin{acknowledgements} 

I am very grateful to R.~Ferlet \& A.~Vidal Madjar for their 
constant help.\\
I would like to thank the referee J.J.~Lissauer for his very 
useful suggestions and comments. 
I also would like to express my gratitude to P.~Barge, H.~Beust, 
A.~Morbidelli and H.~Scholl for many fruitful discussions.
I am particularly indebted to Nissim Kanekar, Ajit Kulkarni and Div. Oberoi
for their critical reading of the manuscript. 

\end{acknowledgements}

\appendix

\section{The carbon bearing gas: a clue to evaporation}
\label{appendix}

In this Appendix we evaluate the total mass of CO present around
$\beta\:$Pictoris\ ($M_{CO}$) and the corresponding supplying rate ($\dot{M_{CO}}$).
We will see that this can give an independent clue to the 
presence of Orbiting-Evaporating-Bodies.

\subsection{The gas around $\beta\:$Pictoris}

The gaseous part of the $\beta\:$Pictoris\ disk 
is detected through the gaseous absorptions superposed 
on the stellar spectrum. These absorptions 
are now classified in four different groups
(see review by Vidal-Madjar et al. 1998):
the interstellar absorption from the local cloud at 10~\kms ; 
a stable component at the stellar velocity (21~\kms ) which is
considered as the gaseous counterpart of the dust disk
(nevertheless the link between the stable gas and the dust
disk is still not established and may even not exist); 
slow and rapid variable absorptions, mainly redshifted, 
and well-explained by the so-called 
Falling-Evaporating-Bodies (FEBs) which are 
star-grazing evaporating comets.
It is thus largely accepted that kilometer-size bodies orbit about $\beta\:$Pictoris\
and some are subject to evaporation 
(see also Lecavelier des Etangs et al. 1997). 

In the FEB scenario, the link between gas and dust is faint, 
there is no direct connection between the presence of dust 
and infalling material. The dust produced simultaneously with 
the gas by Falling-Evaporating-Bodies on eccentric orbits
must be quickly expelled by radiation pressure on hyperbolic orbits.
It is likely that we are now observing two different phenomena
which both take place in the $\beta\:$Pictoris\ system:
presence of dust as around the prototypical star \object{Vega}, 
and presence of falling gas.

\subsection{A needed source of CO and \ci}
\label{CO and ci}

An important characteristic of the $\beta\:$Pictoris\ gaseous disk is
the presence of cold CO and \ci\ (Deleuil et al 1993, Vidal-Madjar et
al. 1994).
Although the CO and \ci\ absorption lines are observed at the stellar 
velocity as the stable component detected by the
absorption lines of single ionized ions 
(\caii , \feii , \mgii , etc...), 
there is evidence that CO and \ci\ have a special status:

- CO is cold with a typical temperature of less than 30~K which corresponds
to the temperature of CO-evaporation; for instance, with an albedo of 0.5 
this temperature corresponds to an evaporating body located between 100 and 200~AU
(Lecavelier des Etangs 1996). This temperature is also consistent
with the observed $^{12}$CO/$^{13}$CO ratio 
of $R=20\pm5$ (Jolly et al. 1998). 
This temperature is obviously very different from the temperature of 
the falling ionized gas: e.g., the \caii\ triplet lines show that 
this gas reaches locally very high temperature 
($T_e> 15\ 000$~K, Mouillet \& Lagrange 1995).

- In contrary to the single ionized ions or \nai , also observed
in the diffuse interstellar
medium, CO and \ci\ are destroyed by UV interstellar photons 
and have lifetime shorter than the star age 
($t_{CO}\sim t_{CI}\sim$200~years). A permanent replenishment mechanism must exist.  

- With the two arguments given above, the supplying
rate of CO (and consequently, \ci ) can be roughly estimated and constrained.
For this estimate, one must assume a cloud geometry which gives
the connection between the observed column density and the total
CO mass. Assuming a disk geometry with an opening angle similar to the dust disk 
($\theta=7\pm$3~degrees), and a characteristic distance 
given by the CO temperature ($r_0=150\pm50$~AU), we get a mass of CO 
\begin{equation}
M_{CO}\approx 4\pi \theta \mu_{CO} N_{CO} r_0^2 \approx 7\times 10^{20} {\rm kg}
\end{equation}
within a factor of 5, where $\mu_{CO}$ is the molecular weight and 
$N_{CO}=2\pm 1 \times 10^{15}$~cm$^{-2}$ is the column density of CO.

Then the known photodissociation rate of CO, $\tau_{CO}=2\cdot 10^{-10}$s$^{-1}$ 
(Van Dishoeck \& Black, 1988)
gives a relation between the total mass and supplying rate.
We obtain
\begin{equation}
\dot{M_{CO}} = M_{CO}\tau_{CO} \approx 10^{11} {\rm kg\ s}^{-1}
\end{equation}

\subsection{Evaporation}

We see that CO and \ci\ probably have a different origin than the 
other observed species and anyway 
need a permanent replenishment mechanism. 
The most obvious mechanism is
certainly the evaporation of cometary-like bodies. In that case, 
CO is ejected from the evaporating body, \ci\ is then produced 
through the CO dissociation (Vidal-Madjar et al. 1994, Jolly et al. 1998).
This evaporation can take place in two different ways:

- In Solar System, frosted bodies like comets are ejected on very 
eccentric orbits 
inside the evaporation zone. The evaporation is thus very rapid, 
these comets become exhausted in a few hundred revolutions.
A lot of volatiles are evaporating: CO, but also CO$_2$ and H$_2$O.

- A second mechanism can also provide evaporation of fros\-ted volatiles: 
the parent bodies can slowly evaporate if their orbit progressively
enters the evaporation limit. In fact, the evaporation rate
is very dependent on the distance to the star ($\sim r^{-20}$, 
Fig.~3 of LVF), 
and the distance below 
which the gas evaporates can thus be considered as very sharp. 
The consequence of such evaporation
of bodies still on quasi-circular orbits is that the dust can remain
on elliptical orbit around the star. For typical grain size distribution
and assuming that the largest grains are larger than 10$\mu$m, 
one can estimate that less than 50\% 
of the mass is ejected on a hyperbolic orbit in the dust tail, 
the remaining is distributed on a large disk structure 
(LVF).
Chiron is subject to such evaporation (Luu \& Jewitt 1990). 
However this cannot happen frequently in the Solar System because the 
evaporation of the common volatiles takes place
inside the planetary system. There, massive planets are responsible for
strong gravitational perturbations and put the evaporating bodies 
on very eccentric orbit; these are then 
observed as classical comets. This is also why Chiron has a chaotic orbit
and will not
remain a slow evaporating body for a long time
(Scholl 1979).

On the contrary, this slow evaporation can occur in an extra-solar planetary 
system if the evaporation limit is outside the massive planets orbits.
For example, this could be the case if a planetary system 
similar to the Solar System was present around a star brighter
than the Sun. 

In LVF, we have estimated the total mass of dust 
associated with the observed CO around $\beta\:$Pictoris, if it is produced in such a 
slow evaporation of orbiting bodies. With reasonable parameters
the mass is found to be consistent with the total mass of observed dust
(see also Sect.~\ref{number of parent bodies});
this provide an {\em independent}
evidence that the $\beta\:$Pictoris\ dust disk can be supplied by 
{\em Orbiting-Evaporating-Bodies}.


\end{document}